\documentclass[apj,twocolumn]{openjournal}
\usepackage{amsmath}
\usepackage{booktabs}
\usepackage{multirow}
\usepackage{color}
\usepackage{soul}

\usepackage[dvipsnames]{xcolor} 

\newcommand{\norm}[1]{\left\lVert#1\right\rVert}

\begin{document}


\title{Predicting Stellar Mass Accretion: An Optimized Echo State Network Approach in
Time Series Modeling} 



\author{Gianfranco Bino$^1$}
\author{Shantanu Basu$^{2,4}$}
\author{Ramit Dey$^{2}$}
\author{Sayantan Auddy$^{5}$}
\author{Lyle Muller$^3$}
\author{Eduard I. Vorobyov$^{6,7}$}
\affiliation{$^1$Department of Applied Mathematics, University of Western Ontario, London, ON, N6A 5B7, Canada}
\affiliation{$^2$Department of Physics \& Astronomy, University of Western Ontario, London, ON, N6A 3K7, Canada}
\affiliation{$^3$Department of Mathematics, University of Western Ontario, London, ON, N6A 5B7, Canada}
\affiliation{$^4$Institute for Earth \& Space Exploration, University of Western Ontario, London, Ontario, N6A 5B7, Canada}
\affiliation{$^5$Jet Propulsion Laboratory, California Institute of Technology, Pasadena, CA 91109, USA}
\affiliation{$^6$Department of Astrophysics, The University of Vienna, A-1180 Vienna, Austria}
\affiliation{$^7$Research Institute of Physics, Southern Federal University, Rostov-on-Don 344090, Russia}




\begin{abstract}
Modeling the dynamics of the formation and evolution of protostellar disks as well as the history of stellar mass accretion typically involve the numerical solution of complex systems of coupled differential equations. The resulting mass accretion history of protostars is known to be highly episodic due to recurrent instabilities and also exhibits short timescale flickering.
By leveraging the strong predictive abilities of neural networks, we extract some of the critical temporal dynamics experienced during the mass accretion including periods of instability. Particularly, we utilize a novel form of the echo state neural network (ESN), which has been shown to deal efficiently with data having inherent nonlinearity. We introduce the use of optimized-ESN (Opt-ESN) to make model-independent time series forecasting of mass accretion rate in the evolution of protostellar disks. We apply the network to multiple hydrodynamic simulations with different initial conditions and exhibiting a variety of temporal dynamics to demonstrate the predictability of the Opt-ESN model. 
The model is trained on simulation data of $\sim 1-2$ Myr, and achieves predictions with a low normalized mean square error ($\sim 10^{-5}$ to $10^{-3}$) for forecasts ranging between 100 and 3800 yr. 
This result shows the promise of the application of machine learning based models to time-domain astronomy.

\keywords{Stellar accretion (1578) --- Neural networks (1933) --- Star formation (1569)}

\end{abstract}

\maketitle 
\section{Introduction}
\label{introduction}
We are entering a new era of rapid advance in time-domain astronomy that promises to revolutionize our understanding of transient astrophysical phenomena. These advances will occur through a variety of instruments that span the electromagnetic and gravitational-wave spectrum. It is important to push forward the development of analysis techniques that can in principle be utilized to model all transient phenomena regardless of the signal source or parameters.

In this paper, we focus on modeling luminosity variations in the evolution of young stellar objects (YSOs). These are precursors of main-sequence stars that form from the collapsing dense regions of interstellar molecular clouds. Stars in their early stages of evolution accumulate materials via mass accretion from the surrounding accretion disk. Mass accretion from the disk to the central object in the early evolutionary stage is likely driven by gravitational torques arising from nonaxisymmetric spiral waves \citep{vor07,vor09}. Other drivers such as disk winds \citep{bai13,suz16}, hydrodynamic and/or magnetohydrodynamic turbulence \citep{bal03}, etc. also lead to redistribution of angular momentum, enabling accretion of disk material onto the central star. However, continuing mass infall onto the disk from larger distances often leads to sustained or recurring gravitational instability (GI) \citep{vor05,vor06} in the disk. The GI further triggers the formation of gas clumps that move inward, resulting in bursts of mass accretion onto the central object.

Such episodic accretion onto YSOs is frequently observed and is well known in the form of FU Orionis objects (FUors) and EX Lupi objects (EXors). Recently, episodic accretion bursts have also been detected in young massive protostars \citep{car17}. The FUors show a rapid rise of luminosity from a few $L_\odot$ to $100-300$ $L_\odot$ \citep{aud14}. This typically corresponds to an increase in mass accretion rate from $10^{-7}$ $M_\odot$ yr$^{-1}$ to a few times $10^{-4}$ $M_\odot$ yr$^{-1}$. The subsequent decline of luminosity after the initial burst occurs over a timescale of many decades. Due to the long timescale of decline, no FUor has ever been observed to have more than one burst. In contrast, the EXors exhibit smaller luminous amplitudes (up to a few tens $L_\odot$) in repetitive outbursts with durations of several months. It is still uncertain whether these two phenomena are related and if FUors correspond to an early stage of evolution with EXors representing smaller amplitude bursts at a later stage \citep[e.g.,][]{con19}.

Furthermore, the mass accretion rates of YSOs are highly episodic due to recurrent instabilities.  They exhibit short timescale flickering due to inherent nonlinearity and inhomogeneity in the disk structure \citep{elb16}. This makes forecasting burst events particularly challenging.
Developing analysis techniques using present-day simulation data is key to advancing the study of such observations even if the underlying dynamics are not well known. 

The last decade has seen phenomenal growth in adaptations of various machine learning (ML) techniques in analyzing astronomical data  \citep{Auddy_2021,auddy2022b} and making predictions in time-domain astronomy \citep{bloom22,rocha22}. Neural network (NN) based models are particularly powerful as they are not tied to a specific set of physical equations and assumptions. They can be trained on data (both from simulation and observation) to capture the nonlinear physics of the system and to make predictions \citep{aud20}. The objective of this paper is to demonstrate that NN-based models can be used to forecast the evolution of transient phenomena in real-time.

We introduce the use of an echo state neural network (ESN) \citep{lukovsevivcius2012practical,kim2020time} to make robust predictions of stellar mass accretion of evolving YSOs. The model is trained on time-series data obtained from hydrodynamical simulations  \citep[see for example][]{vor10}
which capture the evolution of such complex nonlinear star-disk systems. A series of simulations \citep{vor05,vor06,vor10,vor15,mey17} have demonstrated the prevalence of such episodic accretion driven by mass infall onto a nascent protostellar disk. In order to deal with the nonlinearity we use a novel approach of dividing the (simulation) data into a slowly-varying (``deterministic'') component and a more rapidly-varying (``fluctuating'' or ``chaotic'') component. We train the ESN-based model on each component of the data to make the subsequent prediction of the burst events. This ESN-based framework lays the foundation for analyzing such transient phenomena from upcoming surveys, like wide-field optical wavelength mapping with frequent time sampling by the Zwicky Transient Facility (ZTF) and Vera C. Rubin Observatory (VRO). 


This paper is organized as follows. In Section \ref{Hydrodynamic Simulations} we discuss the hydrodynamic simulations that capture the mass accretion in disk evolution. Section \ref{Neural Networks} gives an overview of the ESN architecture. In Section \ref{Optimized Echo State Neural Networks} we introduce the Opt-ESN model and outline the data preparation procedure. Results are presented in Section \ref{results}. A further discussion is in Section \ref{discussion} and conclusions are in Section \ref{conclusion}.
A reader who is focused on the astrophysical consequences of this work can choose to read Section~\ref{Hydrodynamic Simulations} and then move ahead to Section~\ref{results}. 
\section{Hydrodynamic Simulations}
\label{Hydrodynamic Simulations}
Numerical simulations of disk evolution can be done using a set of hydrodynamic equations that are vertically integrated along the direction of the rotation axis and follow the nonaxisymmetric evolution of physical variables in polar ($r,\phi$) coordinates. This is viable in the expected scenario where the disk vertical scale height is significantly less than its radial extent. A series of papers have employed the thin-disk approximation to model the long term evolution of protostellar disks over several Myr timescales \citep[e.g.,][]{vor06,vor10,vor15,vor17,vor20}. It is still challenging to model the full temporal range of disk evolution using three-dimensional simulations, and state-of-the-art models that resolve the central protostar can advance as far as $\sim 10^3$ yr past protostar formation \cite[see][]{mac19}

We train the ESN on the long-term ($\sim 10^6$ yr) disk simulations presented by \cite{vor17}. The simulations calculate the self-consistent disk formation and evolution.  This is done by starting from the hydrodynamic collapse of a prestellar cloud core and continuing into the protostellar phase with a central protostar and surrounding disk. 
The basic equations and numerical finite difference numerical methods are described in \cite{vor10} and \cite{vor17}. A numerical solution is found to the partial differential equations describing the time and space evolution of the mass surface density, the planar momentum components, and the internal energy per unit area. Additional equations are employed to calculate self-gravity, viscosity, and heating and cooling rates due to multiple processes. A central sink cell of radius 5 au is adopted at the coordinate origin in order to avoid very small time steps imposed by the Courant-Friedrichs-Lewy condition, so that the long-term evolution of the remaining region (radius $\sim 10^4$ au) can be calculated.  

The solution of the disk evolution after protostar formation consists of a highly episodic accretion process. While some features of the episodes can be understood in a deterministic manner using the criterion for gravitational instability \citep{das22}, the nonaxisymmetry and nonlinearity of the problem lead to a time evolution of accretion rate that has stochastic and chaotic features.
Each simulation is quite costly in terms of run time (up to several months on a single computer node with 48 physical cores), and a typical parameter survey consists of $\sim 10$ models. What we explore here is the possibility of taking a set of simulation models with different initial conditions as input and training a neural network on a portion of the time evolution in order to extract some intrinsic and underlying dynamics of the system. We can then see how far in time the neural network can forecast the solution into a regime where it was not trained. 



\subsection{Hydrodynamic Simulation Outputs}
\label{Hydrodynamic Simulation Outputs}
We utilize the simulation outputs in 6 of the 35 models presented in \cite{vor17}. 
These six models differ in their initial conditions, which are described here.
The initial axisymmetric radial profiles for the gas surface density $\Sigma$ and angular velocity $\Omega$ for the initial prestellar collapsing core are
\begin{align}
    \Sigma &= \frac{r_0 \Sigma_0}{\sqrt{r^2 + r_0^2}}\, , \\
    \Omega &= 2\, \Omega_0 \left( \frac{r_0}{r} \right)^2 \left[ \sqrt{1 + \left( \frac{r}{r_0} \right)^2} - 1 \right]\, ,
\end{align}
where $\Sigma_0$ and $\Omega_0$ are the surface mass density and angular velocity at the center of the core, respectively. These are power-law profiles with asymptotic dependence $\propto r^{-1}$ and have a central plateau radius 
$r_0$ that is the length scale over which thermal pressure can smoothen the density profile \citep[for details, see][]{vor17}. 

To generate a gravitationally unstable core, each model is characterized by the ratio $r_{\text{out}}/r_0 = 6$, where $r_{\text{out}}$ is the core's outer radius. 
The cloud core mass $M_{\text{cl}}$ is found using the initial radial profile for the gas surface density $\Sigma$. The quantity $\Omega_0$ is selected such that the models have an intial ratio of rotational to gravitational energy $\beta_0$ in the range of $\approx 10^{-4}$ to $0.07$. We summarize the initial model conditions in Table \ref{tab::m_acc}. 

\begin{figure*}
\hspace*{-0.8cm}
\centering
\includegraphics[width=0.95\textwidth]{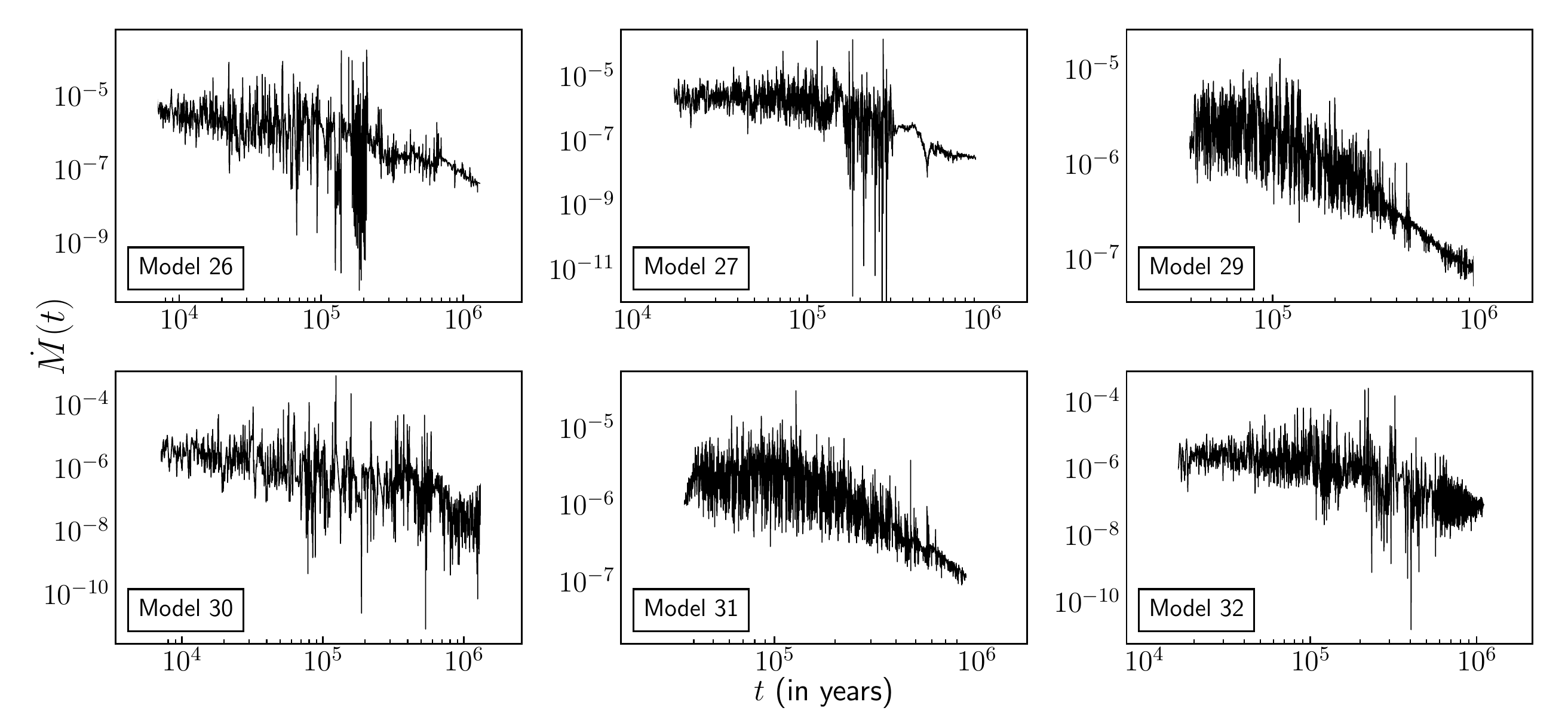}
\vspace*{-0.3cm}
  \caption{Mass accretion rate evolution from hydrodynamic simulations for each of the six models labeled 26 through 32 (with the exception of model 28) in \cite{vor17}. The mass accretion rate is shown in units of $M_\odot$ yr$^{-1}$. }
  \label{sim_data}               
\end{figure*}

The simulation outputs of the mass accretion rate to the central sink $\dot{M}(t)$ are shown in Figure \ref{sim_data} for each model. Comparison with the Table \ref{tab::m_acc} parameter values shows that there is a general increase of the variability amplitude as $M_{\rm core}$ and/or $\beta_0$ increase. Increasing mass or angular momentum leads to more massive protostellar disks and greater activity of GI induced bursts.
\section{Echo state neural networks}
\label{Neural Networks}
\begin{figure}
\centering
  \includegraphics[width=0.45\textwidth]{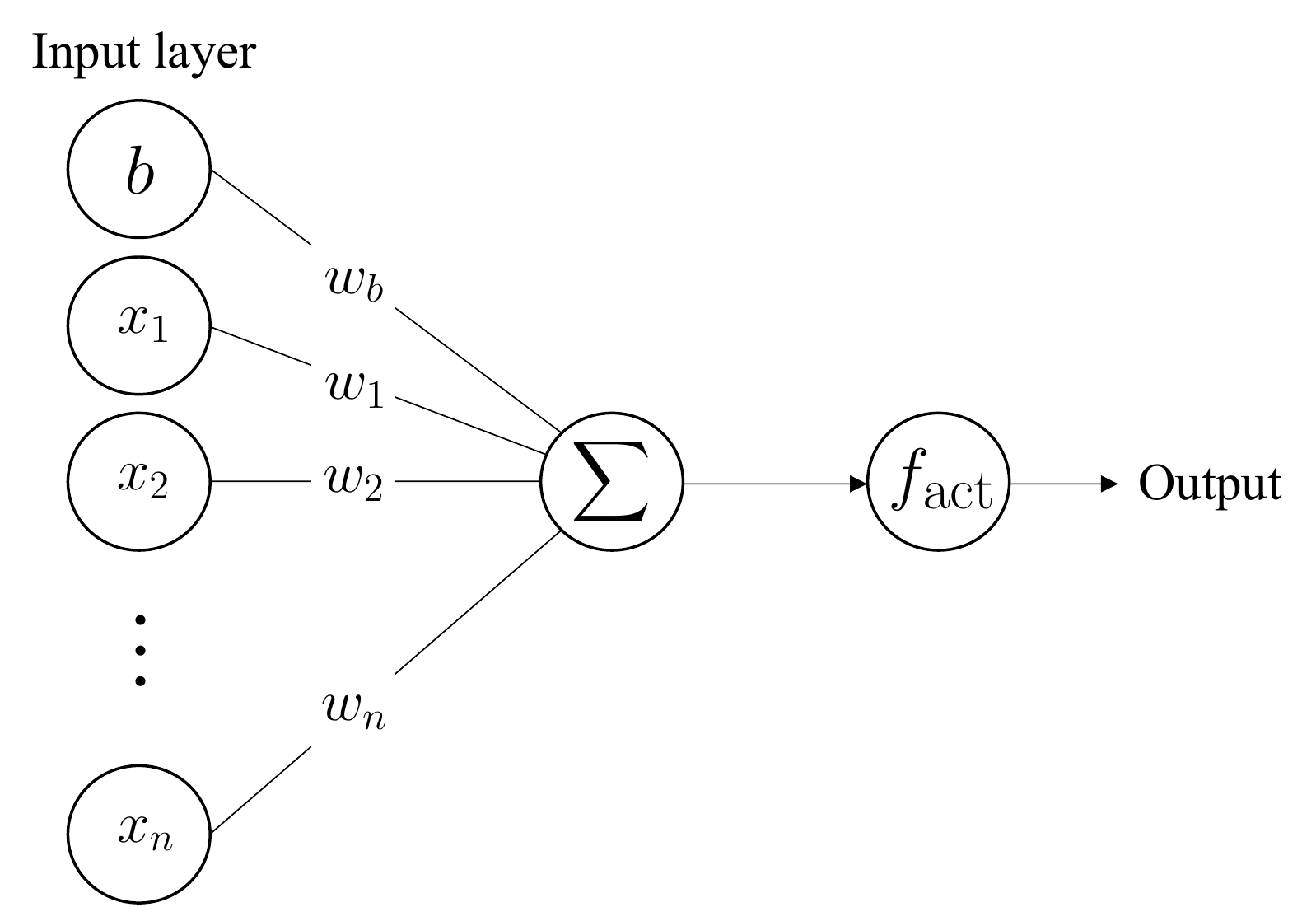}
  \caption{Schematic diagram of a single layer perceptron. The input data and bias are given as $\textbf{x} = [b, x_1, x_2, \ldots , x_n]$ while the weights are given as  $\textbf{W} = [w_b, w_1, w_2, \ldots , w_n]$. $f_{\text{act}}$ is the activation function through which $ \sum_i W_i x_i$ is passed to return an output $y=f_{act}\left( \sum_i W_i x_i \right)$. }
  \label{neural_net1}
\end{figure}

\begin{figure}
\centering
   \includegraphics[width=0.45\textwidth]{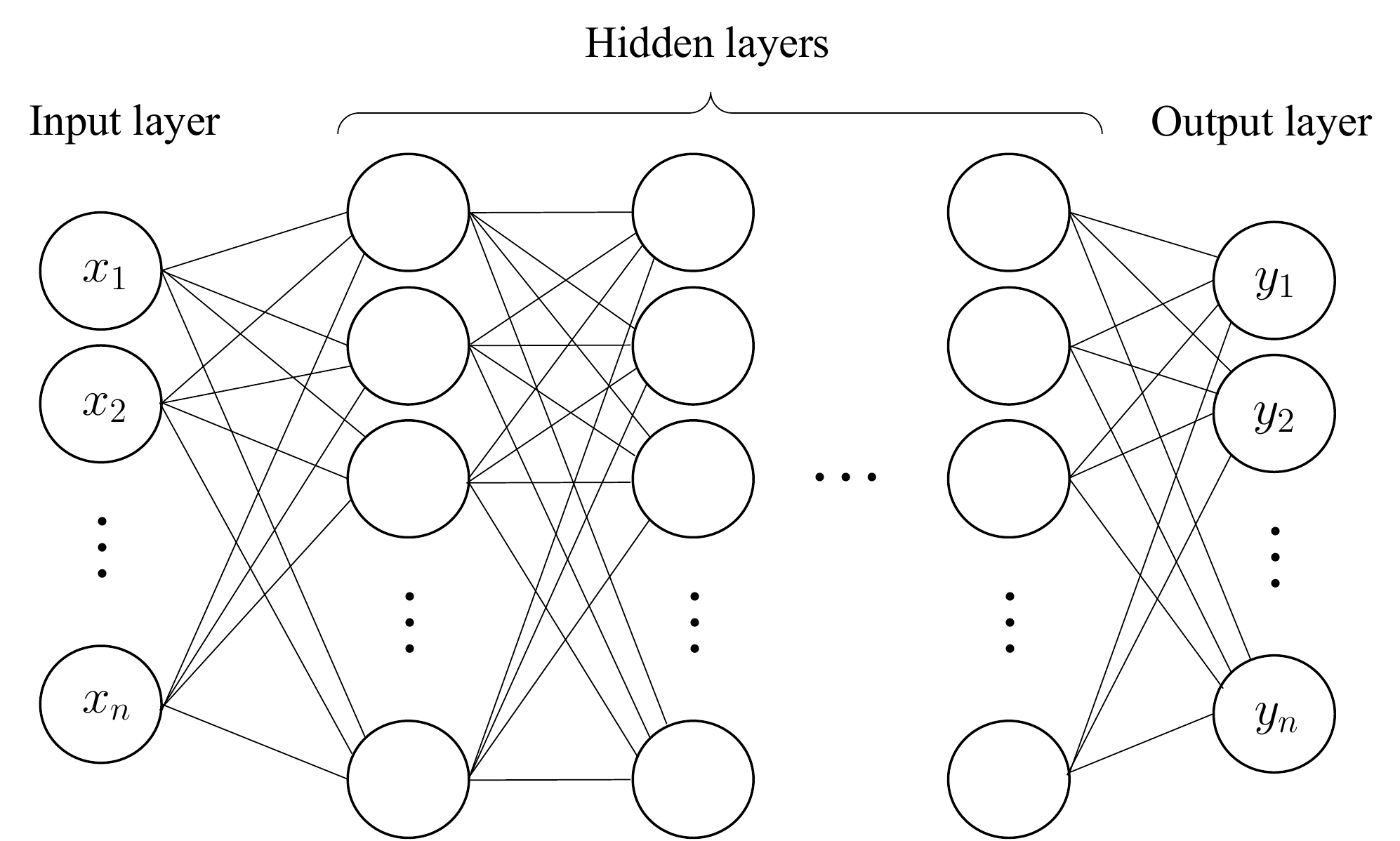}
   \vspace{0.2cm}
  \caption{Schematic diagram of a multilayer perceptron neural network with several hidden layers. The neurons in the hidden layers take a linear combination of the inputs from the previous layer and passes it through an activation function to generate an output, which is further passed to the set of neurons in the next layer.}
  \label{neural_net2}
\end{figure}

Neural networks (NN) have demonstrated the capability to approximate continuous functions and are often referred to as universal approximators \citep{schafer2006recurrent}. In the context of time series analysis, this gives them the ability to estimate the underlying dynamical processes governing the system. This is done by using the NN as a mapping function between the inputs and targeted outputs, allowing them to extract complex temporal relationships within the time-series data. The NN architecture is based on a collection of interconnected nodes, or ``neurons'', as shown schematically in Figure~\ref{neural_net1}. The nodes are often arranged in layers from input to output, as shown in Figure \ref{neural_net2}. These architectures can be further extended to include recurrent units that maintain a network's hidden state during model training. These hidden states allow the network to recognize temporal sequences in data, which can make a recurrent neural network (RNN) \citep{2018arXiv180101078S} particularly useful in the context of time series analysis.

%
%
\begin{figure}
\centering
  \includegraphics[width=0.45\textwidth]{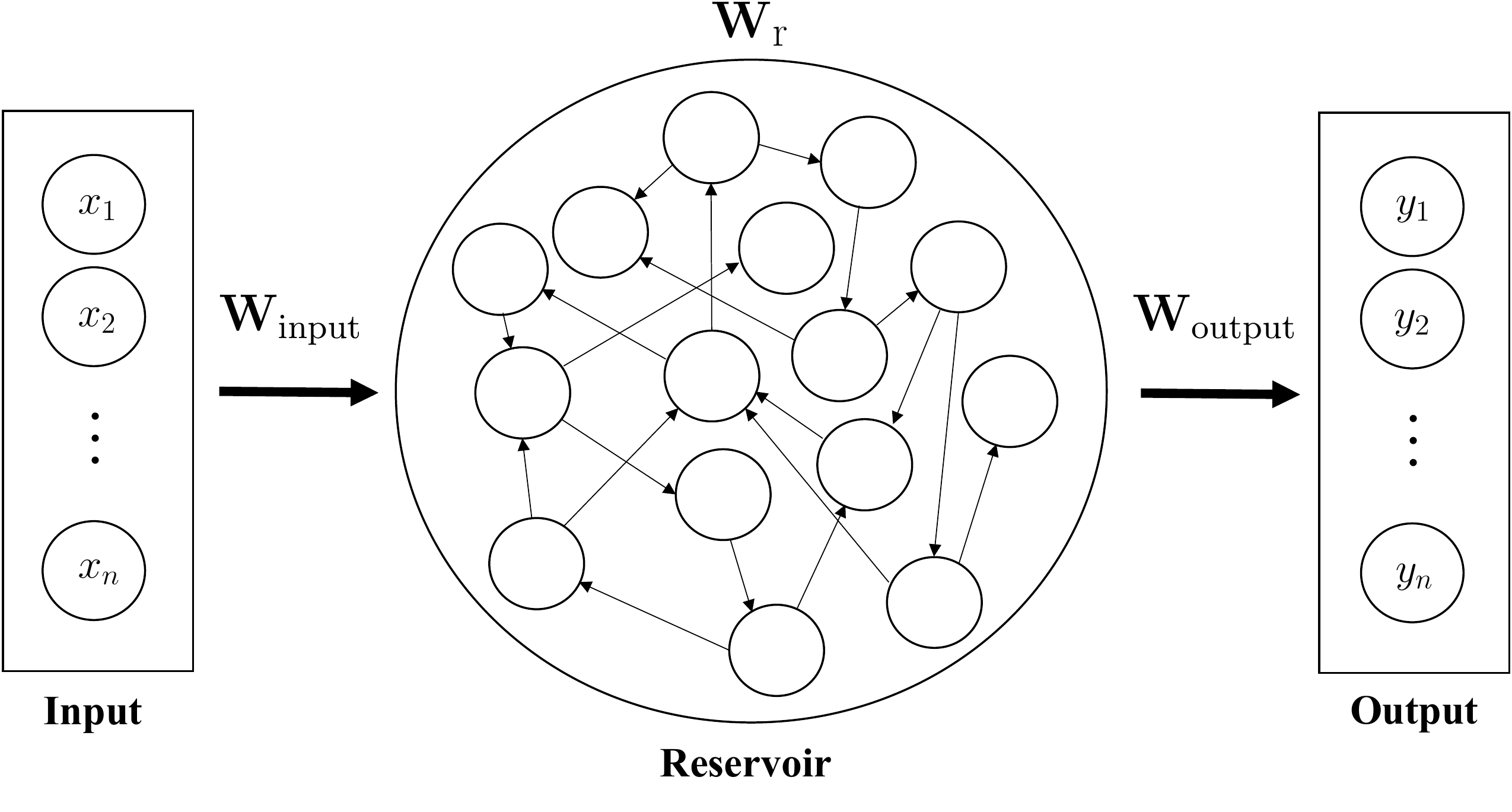}
  \caption{General architecture of an echo state neural network, shown with sparsely connected random reservoir units. Here, $\textbf{W}_{\text{input}}$ and $\textbf{W}_{r}$ are randomly generated sparse matrices.  }
  \vspace{0.1cm}
  \label{neural_net3}
\end{figure}
\begin{figure*}
\centering
  \includegraphics[width=0.8\textwidth]{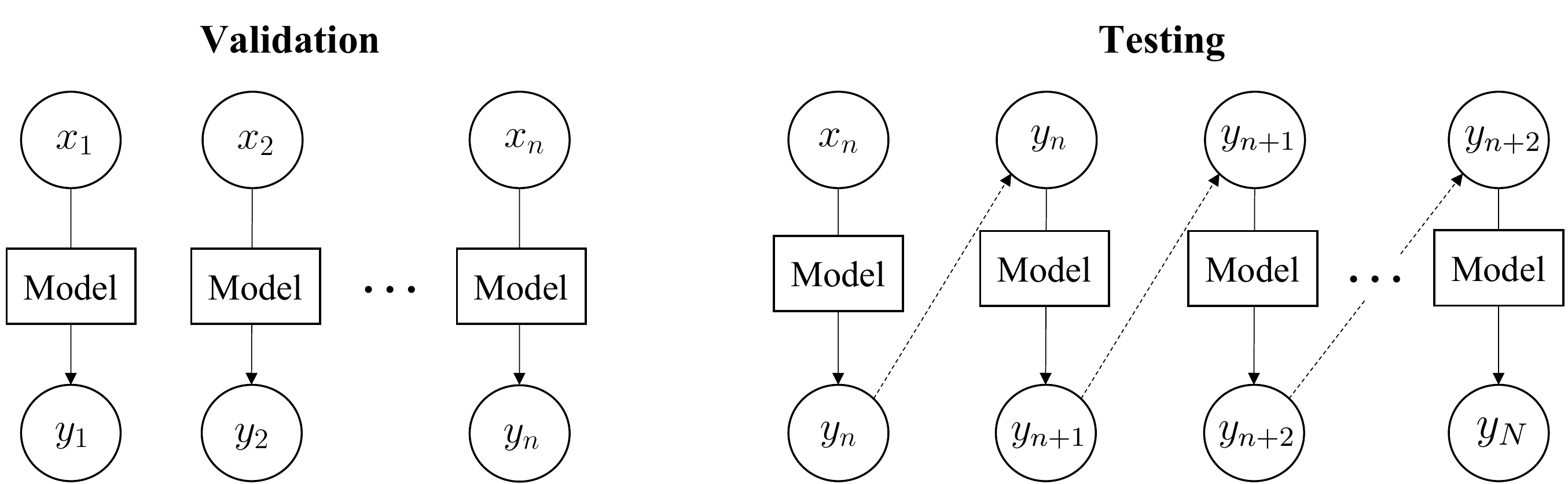}
  \vspace{0.2cm}
  \caption{Input to output sequence for both the validation (left) and testing (right) sets. The validation set is considered ``in-time'' because each output value $y_i$ is calculated using an input $x_i$ taken directly from the given data set. The testing set however, is considered ``out-of-time'' because the model output is recursively served as an input to the following time step. That is, given the data set's terminal point $x_n$, the prediction of $x(t)$ at the following point in time is given by the network output $y_n$. As we step beyond the horizon of the given data set, $y_n$ will be used as the input to calculate the prediction $y_{n+1}$ and so on. } 
  \label{val_test}
\end{figure*}
An ESN can be considered to be a sparsely connected RNN where the hidden layers along with the weights act as a ``reservoir". 
This reservoir functions as a nonlinear temporal kernel, embedding the dynamics of the input data onto a higher dimensional computation space. For an ESN architecture only the reservoir-to-output weights are trainable while the input-to-hidden and hidden-to-hidden weights are chosen randomly and kept fixed during the training process. The sparsity of the ESN architecture and the fact that the hidden layer weights are not updated during the training process, automatically addresses the problem of vanishing gradients, as seen typically for a more conventional RNN based model \citep{jaeger2007echo,lukovsevivcius2012practical}. Since only the output weights are trainable, which is a simple linear regression task compared to the slow convergence of tuning the parameters of other networks, ESNs are much faster to train compared to other RNNs. 
For chaotic time-series prediction, ESNs have shown exceptionally good performance as these networks can capture the nonlinear dynamics of the system efficiently.  

In order to effectively model the chaotic dynamics that govern the hydrodynamic simulations, we implement the use of an ESN based model. An ESN architecture having a input, reservoir and output layer with its corresponding weights is demonstrated in Figure \ref{neural_net3}. For an input time series $\textbf{x}(t)$, we begin by defining the following input and target time sequences:
\begin{flalign*}
&\textbf{x}_1(t) = [x_1(t), x_2(t), \cdots, x_{n-1}(t)] \quad \; \text{Input Sequence} \\
&\textbf{x}_2(t) = [x_2(t), x_3(t), \cdots, x_n(t)] \quad \quad \; \text{Target Sequence}
\end{flalign*}
In order to utilize the neural network $\mathcal{N}$ as a predictive time series model, we form the mapping $\mathcal{N} : \textbf{x}_1(t) \mapsto \textbf{x}_2(t)$ to extract the relationship between the quantities $x_i(t)$ and $x_{i+1}(t)$. To do so, we train the ESN using the following steps:
\begin{itemize}
    \item Randomly generate the the input weight matrix $\textbf{W}_{\text{input}}$ and the reservoir weight matrix $\textbf{W}_{r}$.
    \item For each quantity $x_i(t)$ in $\textbf{x}_1(t)$, construct an $N_r \times 1$ reservoir state vector $\textbf{v}_i$, initialized to $\textbf{v}_1 = \textbf{0}$. Let
\begin{align}
&\textbf{v}_{i+1} = (1- \alpha) \textbf{v}_i \label{res_state} + \alpha f_{\mathrm{act}}( \mathcal{W}_i ) \, , \\
&\text{with} \quad \mathcal{W}_i = \textbf{W}_{\mathrm{input}} x_i(t) + \textbf{W}_r \textbf{v}_i + \textbf{W}_{b}\, , \nonumber
\end{align}
    where $N_r$ is the reservoir size and $0 < \alpha < 1$ is the leaking rate. Equation (\ref{res_state}) includes a randomly generated bias term $\textbf{W}_{b}$ and adopts the activation function $f_{\mathrm{act}}(x) = \tanh(x)$.
    \item Define a washout quantity $\omega < n$ as an initially discarded transient and for every $i > \omega$, construct the internal state 
\begin{equation}
\textbf{X} = \begin{bmatrix}
   1  & 1 & \cdots & 1\\
   x_i (t) & x_{i+1}(t)  & \cdots & x_{n-1}(t) \\
   \textbf{v}_i & \textbf{v}_{i+1} & \cdots & \textbf{v}_{n-1} \\
\end{bmatrix}.
\end{equation}
\item Finally, compute the output matrix using the Moore-Penrose inverse on the set $\overline{\textbf{x}}_2(t) = [x_{\omega +2}(t), x_{\omega + 3}(t), \cdots, x_n(t)]$, yielding
\begin{equation}
\textbf{W}_{\mathrm{output}} = \overline{\textbf{x}}_2^{\ast} \left( \textbf{X}^{\ast} \textbf{X} \right) ^{-1} \textbf{X}^{\ast}.
\end{equation}
Note that if the matrix $\left( \textbf{X}^{\ast} \textbf{X} \right)$ is near-singular, it is recommended to regularize the regression by adding a constant $\lambda$ along the diagonals (known as Tikhonov regularization).
\end{itemize}
Once $\textbf{W}_{\mathrm{output}}$ has been calculated, the output can be computed as
\begin{equation}
\textbf{y}(t) = \textbf{W}_{\mathrm{output}} \textbf{X} \label{outY}\, .
\end{equation}
This effectively represents an estimate to the mapping of one point in time to the next. Therefore, in order to predict the $(i+1)$th time step (i.e., estimate the quantity $x_{i+1}$), we construct the internal state \textbf{v} with $x_i$ and use $\textbf{W}_{\mathrm{output}}$ to compute the output using Equation (\ref{outY}).\\

The hyperparameters used for defining the reservoir and characterzing the network are described below:

\begin{itemize}

    \item The reservoir size $N_r$\\
    Determines the number of units in the reservoir (or in turn the size of the reservoir).
    
    \item Spectral Radius $\rho$ \\
    This is a global parameter that determines the maximal eigenvalue of the $\textbf{W}_{r}$ matrix. In other words it scales the reservoir connection matrix and controls the width of the distribution of the nonzero elements present in  $\textbf{W}_{r}$. In most cases, $\rho(\textbf{W}) < 1$ maintains the echo state property. 
    \item Input scaling $\varrho$ \\
    This parameter determines the scaling of the input weight matrix. It also controls the amount of nonlinearity in the dynamics of the reservoir.
    \item Connectivity $c_r$ \\
    Controls the degree of sparsity in the reservoir weight matrix.
    \item Leaking Rate $\alpha$ \\
    Controls the speed of the reservoir dynamics in reaction to the input.
\end{itemize}

\section{Optimized Echo State Neural Networks}
\label{Optimized Echo State Neural Networks}

\subsection{Network Architecture}
\cite{liu2018} introduced a parallel series approach where one stacks a series of reservoirs by generating $L$ independent input and reservoir matrices. The time series is trained and validated through each reservoir to form $L$ output matrices, 
Finally, the model’s output $\hat{\textbf{y}}(t)$ is taken to be the mean of all $L$ realizations so that
\begin{equation}
    \hat{\textbf{y}}(t) = \frac{1}{L} \sum_j \textbf{y}^{(j)}(t)\, . \label{eq10}
\end{equation}
The optimized-ESN (Opt-ESN) extends Equation (\ref{eq10}) to being a weighted sum rather than the standard mean. That is, the output realizations from each reservoir are weighted by a set of optimal coefficients. As such, the final output of the Opt-ESN is given by
\begin{equation}
    \hat{\textbf{y}}(t; \hat{\boldsymbol \beta}) = \sum_j \hat{\beta}_j \textbf{y}^{(j)}(t) \, ,
\end{equation}
where the coefficients $\hat{\boldsymbol \beta}$ are found by minimizing the squared residuals over the input's validation segment $\textbf{x}_{\text{val}}(t)$. This is done by solving the linear optimization problem to find the minimum value of the loss function
\begin{equation}
\begin{array}{rl}
\mathcal{L} = \norm{\textbf{x}_{\text{val}}(t) - \hat{\textbf{y}}(t; \hat{\boldsymbol \beta})} ^2

\end{array}. \label{opt}
\end{equation}
Here, the validation segment is defined as the in-time portion of the data set used to validate the model's output (see Section \ref{Data segmentation} for more details). 

\subsection{Data Preparation}
\label{Data preparation}
\begin{figure*}[h]
\centering
  \includegraphics[width=\textwidth]{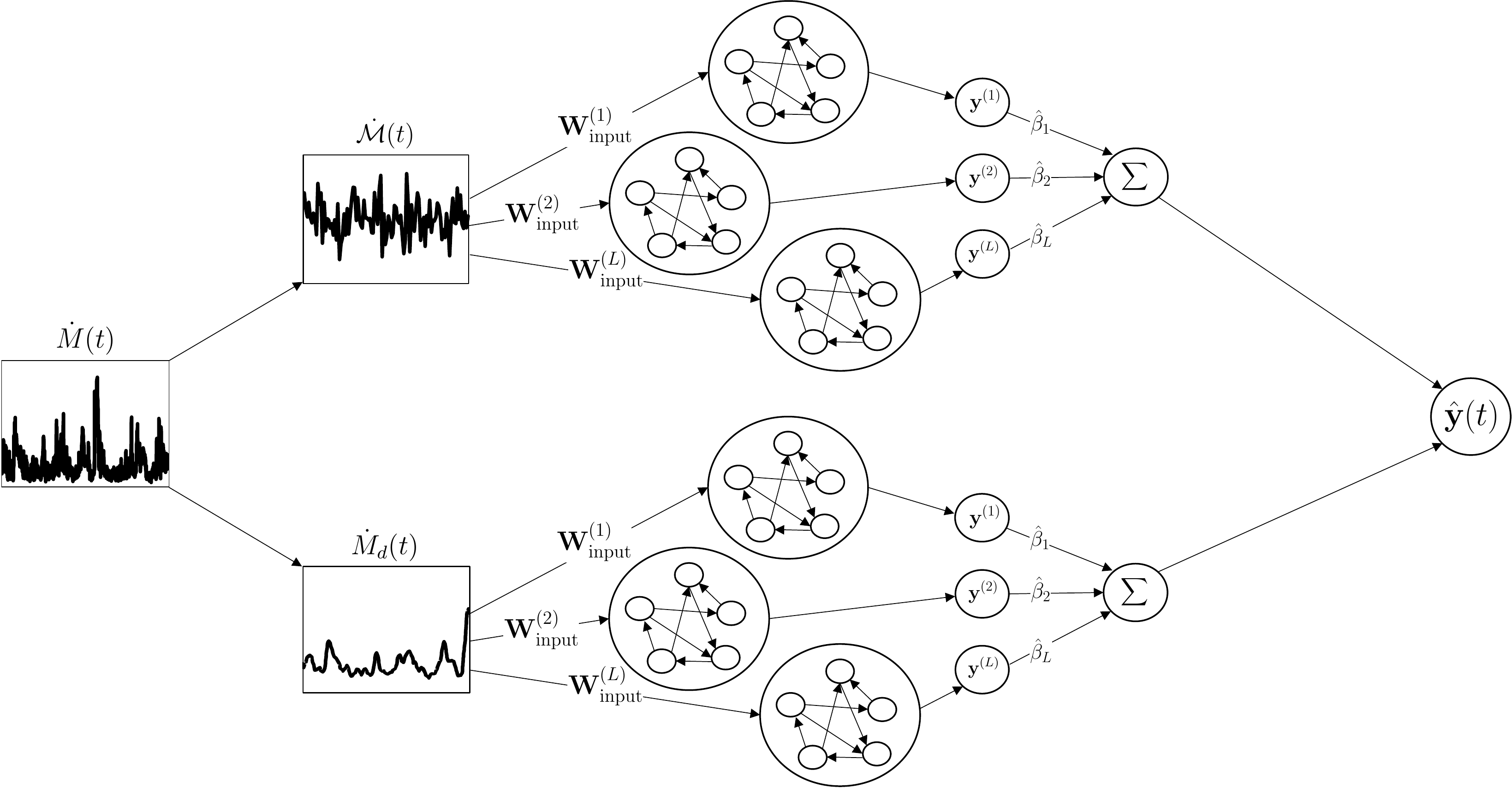}
  \vspace{0.15cm}
  \caption{Schematic figure representing the proposed architecture of the Opt-ESN. The input is separated into fluctuating and deterministic components and fed through an Opt-ESN. Each of the individual forecasts are summed to produce the estimated forecast to $\dot{M}(t)$. Here, the neural network is shown to contain $L$ reservoirs stacked into a single layer used in formulating the linear programming problem. Each reservoir is taken to be sparse and randomly connected.}
  \label{neural_net4}
\end{figure*}

As the input to the Opt-ESN, we used a portion of the simulated data for each model, corresponding to times of vigorous episodic accretion.
We divide the data into segments of various lengths having different time steps and use this for training, validation and testing. Furthermore, as an aid to assessing the quality of our forecasts, we characterize the time scale in terms of the Lyapunov exponent. We define the dimensionless time length $\Lambda \cdot N_t$ as the Lyapunov time, where $N_t$ is the observation number. Here, the quantity $\Lambda$ is taken to be the maximum Lyapunov exponent which characterizes the rate of separation between close trajectories in phase space and effectively quantifies the degree of chaos present. That is, if two trajectories are initially separated by some infinitesimal amount $\Delta_0$, then the rate of divergence as a function of time $t$ is approximately 
\begin{align}
    | \Delta (t)  | \approx | \Delta_0 | \cdot \text{exp}(\Lambda t).
\end{align}
It becomes clear that for $\Lambda > 0$, the separation $| \Delta (t)  | $ grows exponentially with time \citep{vulpiani2009chaos}. We estimate $\Lambda$ using the algorithm in \cite{eckmann1986liapunov} where our outputs are given in the fourth column of Table \ref{tab::per_summ}. Our calculations demonstrate that each value of $\Lambda$ is greater than zero, indicating that each of the simulation models can be considered to be a chaotic system. We preprocess the simulation data $\dot{M}(t)$ by assuming that it is separable in the form
\begin{equation}
\dot{M}(t) = \dot{M}_d(t) + \dot{\mathcal{M}}(t)\, ,
\end{equation}
where $\dot{M}_d(t)$ and $\dot{\mathcal{M}}(t)$ represent the data's deterministic and fluctuating components, respectively. In order to extract the fluctuating component, we pass $\dot{M}(t)$ through a high pass filter\footnote{We utilized Matlab's \texttt{highpass} function with passband frequency \texttt{fpass = 2000} and sampling rate \texttt{f = 50000}. The units are the inverse observation number.}. The deterministic component can then be extracted by subtracting $\dot{\mathcal{M}}(t)$ from $\dot{M}(t)$. Furthermore, we normalize each component with respect to the standard deviation $\sigma$ of $\dot{M}(t)$. In order to make predictions, we feed $\dot{M}_d(t)$ and $\dot{\mathcal{M}}(t)$ into the Opt-ESN separately, run the forecasts and sum the outputs to get the final prediction on $\dot{M}(t)$. We found that decomposing and processing the simulation data in this fashion gives better performing output than inputting $\dot{M}(t)$ directly. The proposed network architecture is given in Figure \ref{neural_net4}. The network has two important layers contributing to the output. At the stacked reservoir layer, the output weight matrices $\textbf{W}_{\text{output}}$ are determined using regression techniques over the training segment. The weight matrices are then used to output a series of unique paths over the validation segment where the coefficients $\hat{\beta}_j$ are found by solving a linear optimization problem.

\subsection{Hyperparameter Selection}
\label{Hyperparameter Tuning}
Selecting the optimal set of hyperparameters can be difficult given the size of the hyperparameter space. As such, for each component of the individual simulation data set (fluctuating and deterministic), we perform a hybrid, discrete-stochastic search over the entire hyperparameter space. That is, we prespecify the search space for each hyperparameter, construct a nested loop searching over all possible combinations, build a model and compute the mean square error (MSE) on the validation segment. In addition to this, within each iteration of the loop, we generate a random number for each hyperparameter that is within the search space, construct and validate a second model, and compute the respective MSE. These two models are then compared and the one with the lowest MSE is kept. This process continues until the entire hyperparameter space has been searched and the model with the lowest MSE on the validation segment is selected.
\begin{figure*}[h]
\centering
  \includegraphics[width=0.89\textwidth]{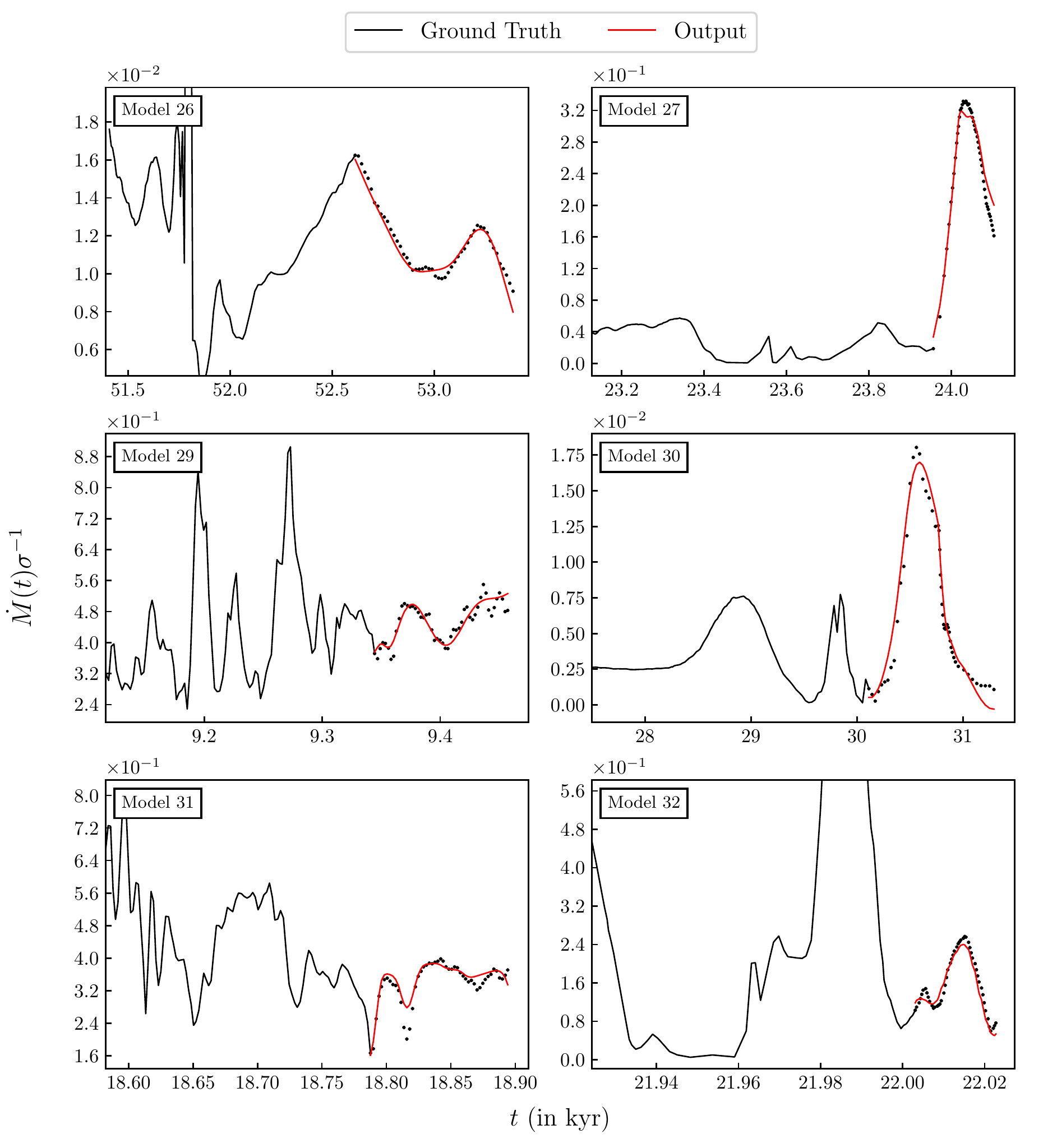}
  \caption{The Opt-ESN predictions of the simulation data. The solid black line shows a fraction of the simulation data used for training/validation, the dotted black is the out-of-time testing segment of the data and the red line is the network's prediction. 
  } \label{modOut}
\end{figure*}
\begin{figure*}
\centering
  \includegraphics[width=\textwidth]{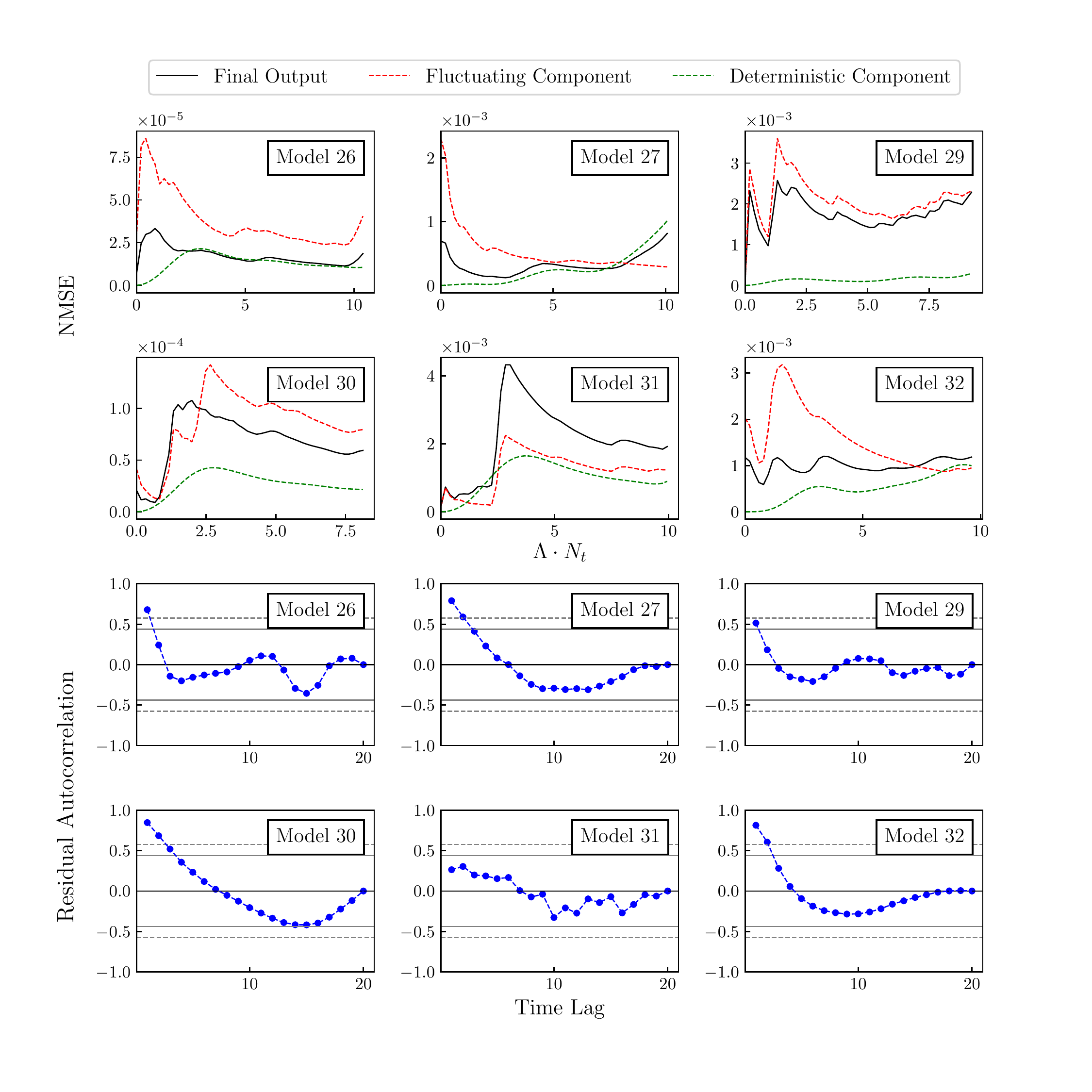}
  \vspace{-1.2cm}
  \caption{Performance assessments on the predictions made for the simulation data. The upper portion of the figure demonstrates the rolling NMSE as a function of Lyapunov time for $\dot{M}(t)$, $\dot{\mathcal{M}}(t)$ and  $\dot{M}_d(t)$. The lower portion of the figure gives the residual autocorrelation function on the validation segment.} \label{modgof}
\end{figure*} 

\subsubsection{Training and Data Segmentation}
\label{Data segmentation}
In order to train the Opt-ESN model, the input data must be split into three segments:
\begin{itemize}
    \item Training Segment ($t_0 < t \leq t_{\text{train}}$)\\ 
    Used to train the model and compute the output matrix $\textbf{W}_{\mathrm{output}}$.
    \item Validation Segment ($t_{\text{train}} < t \leq t_{\text{val}}$)\\
    Used as an in-time assessment on how well $\textbf{W}_{\mathrm{output}}$ maps a single input to an output.
    \item Testing Segment ($t_{\text{val}} < t \leq T$) \\
    Used as a blind test for the model to assess how well the out-of-time predictions perform.
\end{itemize}
The difference between the validation and testing segment is that the model is dynamic within the validation segment. That is, each output is calculated using an input that is directly from the validation segment of the data. In the testing segment however, the model is using it's own output as an input for the next time step. This is demonstrated in Figure \ref{val_test} where the validation use is given on the left side diagram and the testing use is given on the right side diagram.
\begin{table}
\vspace{0.25cm}
\begin{center}
\caption{Initial conditions for each of the seven simulation models.}
\begin{tabular}{c c c c c}
\hline \hline
{\textbf{Model}} & {$M_{\text{core}}$} & {$\beta_0$} & {$r_0$} & {$M_{*, \text{fin}}$}\\
\hline 
Model 26 &  $1.245 M_{\odot}$ & $1.27 \%$ & $2777$ au & $0.753M_{\odot}$\\
Model 27 &  $1.076 M_{\odot}$ & $0.56 \%$ & $2400$ au & $0.801M_{\odot}$\\
Model 29 &  $0.999 M_{\odot}$ & $0.28 \%$ & $2229$ au & $0.818M_{\odot}$\\
Model 30 &  $1.537 M_{\odot}$ & $1.27 \%$ & $3429$ au & $0.887M_{\odot}$\\
Model 31 &  $1.306 M_{\odot}$ & $0.28 \%$ & $2915$ au & $1.031M_{\odot}$\\
Model 32 &  $1.383 M_{\odot}$ & $0.56 \%$ & $3086$ au & $1.070M_{\odot}$\\
\hline \hline
\end{tabular}
\label{tab::m_acc}
\end{center}
\end{table}

\section{Results}
\label{results}

\subsection{Opt-ESN Outputs}
\label{Model Outputs}
In order to train the model, the simulation data for each model is decomposed into deterministic and fluctuating components and standardized as discussed in Section \ref{Data preparation}. Forecasting either component of the simulation data requires the time series to be stationary. Nonstationary dynamics risk consequences such as spurious correlations and heavily biased mean and variance estimates (see Appendix for more details). We assess the stationarity of each component using the augmented Dickey–Fuller (ADF) test at 5\% significance \citep{patterson2011unit}. Ultimately, we find that in each model, $\dot{\mathcal{M}}(t)$ follows a stationary process while $\dot{M}_d(t)$ does not. As such, we applied first order differencing on the deterministic component to enforce stationarity.
Network hyperparameters are selected according to the methodology introduced in Section \ref{Hyperparameter Tuning}. We utilize a stack of $L = 10$ reservoirs for size $N_r = 250$ neurons and solve the optimization problem for Equation (\ref{opt}) to generate the final output. In Figure \ref{modOut}, the Opt-ESN forecast is given for all six simulation models. The outputs are found by summing the individual forecasts of $\dot{M}_d(t)$ and $\dot{\mathcal{M}}(t)$ (the individual forecasts are shown in the Appendix). 

We assess the Opt-ESN performance using the dimensionless normalized mean square error
\begin{align}
    \text{NMSE} = \frac{1}{n} \sum_{i=1}^n \frac{(y_i - \hat{y_i})^2}{\text{max}(\hat{y}) - \text{min}(\hat{y}) }\, ,
 \end{align}
where $y_i$ and $\hat{y_i}$ are the observed and predicted values, respectively, for $n$ data points. Here, we take $y = \dot{M} / \sigma$, which is a dimensionless quantity. Lower NMSE values indicate stronger performance and NMSE $= 0$ equates to perfect accuracy. We consider any model having NMSE $< 10^{-2}$ to indicate good performance. The performance assessments are summarized in Table \ref{tab::per_summ} in the Appendix. Additionally, in the upper half of Figure \ref{modgof}, the rolling NMSE on each model component is given as a function of Lyapunov time. The Lyapunov time is the characteristic timescale on which a system is chaotic and typically limits the predictability of the said system. Here, the Lyapunov time is calculated using the maximum Lyapunov exponent, $\Lambda = \Lambda_{\textrm{max}}$. If the model is well specified, the residuals between the observed and predicted values are expected to be approximately normally distributed and thus largely attributed to white noise.

\begin{figure*}
\centering
  \includegraphics[width=\textwidth]{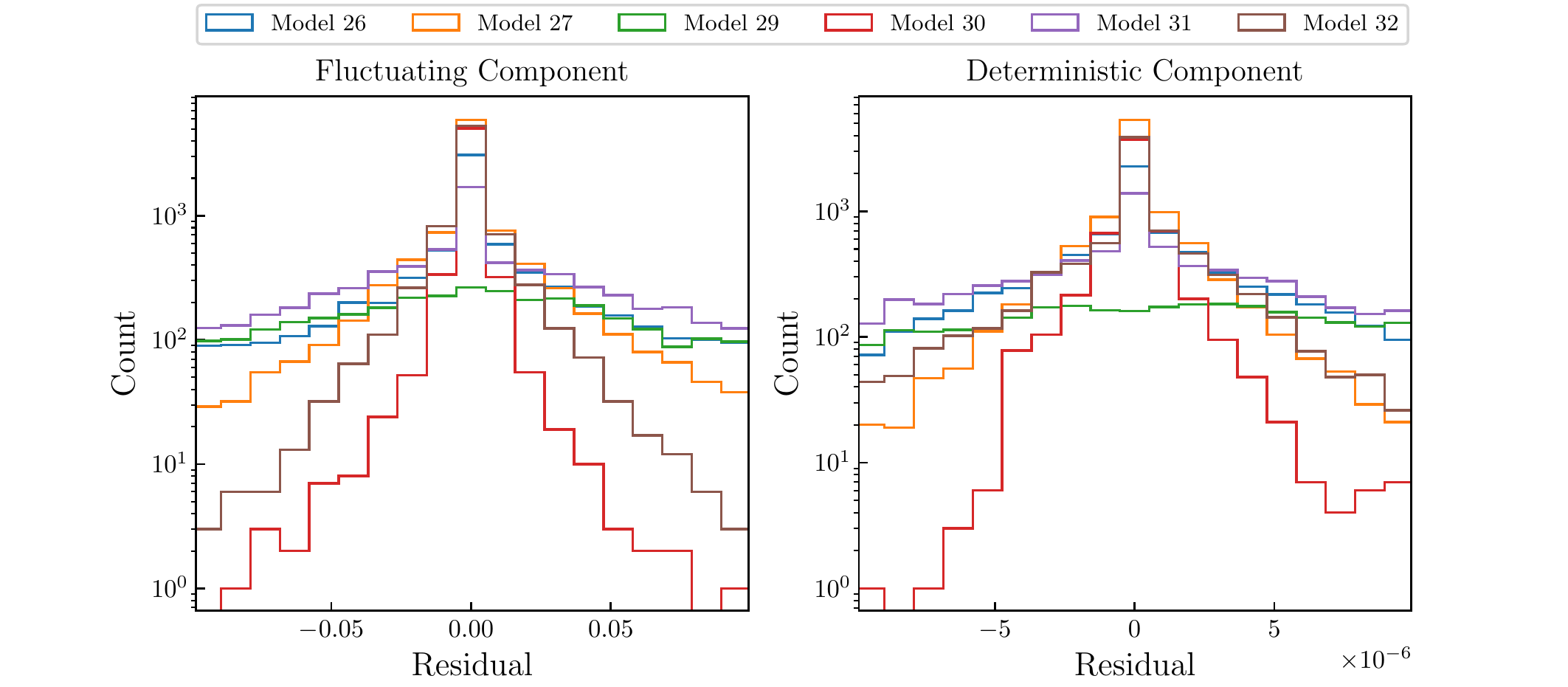}
  \vspace{-0.4cm}
  \caption{Distribution of residuals for each simulation model after network training. Goodness-of-fit assessed using the one-sample $t$-test for the null hypothesis that the residuals are sampled from a normal distribution with mean equal to zero.} \label{modres}
\end{figure*}

Over the validation set, our goodness-of-fit assessment on the residuals utilize the autocorrelation function. In the lower portion of Figure \ref{modgof}, the autocorrelations are given for lagged residuals adopting a 99\% and 95\% confidence interval represented by the solid boundary lines and dashed boundary lines, respectively. Given that at least 90\% of lagged residuals lie within the confidence interval, we assume the process to be approximately white noise. Furthermore, the residuals over the training phase are used to assess the goodness-of-fit of the Opt-ESN. We achieve this using the  one-sample $t$-test for the null hypothesis that the residuals are sampled from a normal distribution with zero mean. The test statistic is
$$
t^* = \frac{\bar{x} - \mu}{\sigma_s/\sqrt{n}} \, ,
$$
where $\bar{x}$ is the sample mean, $\mu$ is the hypothesized mean, $\sigma_s$ is the sample standard deviation and $n$ is the sample size. The statistic follows a $t$-distribution with $n-1$ degrees of freedom. Our results show that for each model, we cannot reject the null hypothesis at the 5\% significance, indicating that the residuals are approximately normal with zero mean. Figure \ref{modres} shows the residual distributions for each model.
 \subsubsection{Episodic Bursts and Stability Assessment}\label{burstsec}
\begin{figure*}
\centering
  \includegraphics[width=\textwidth]{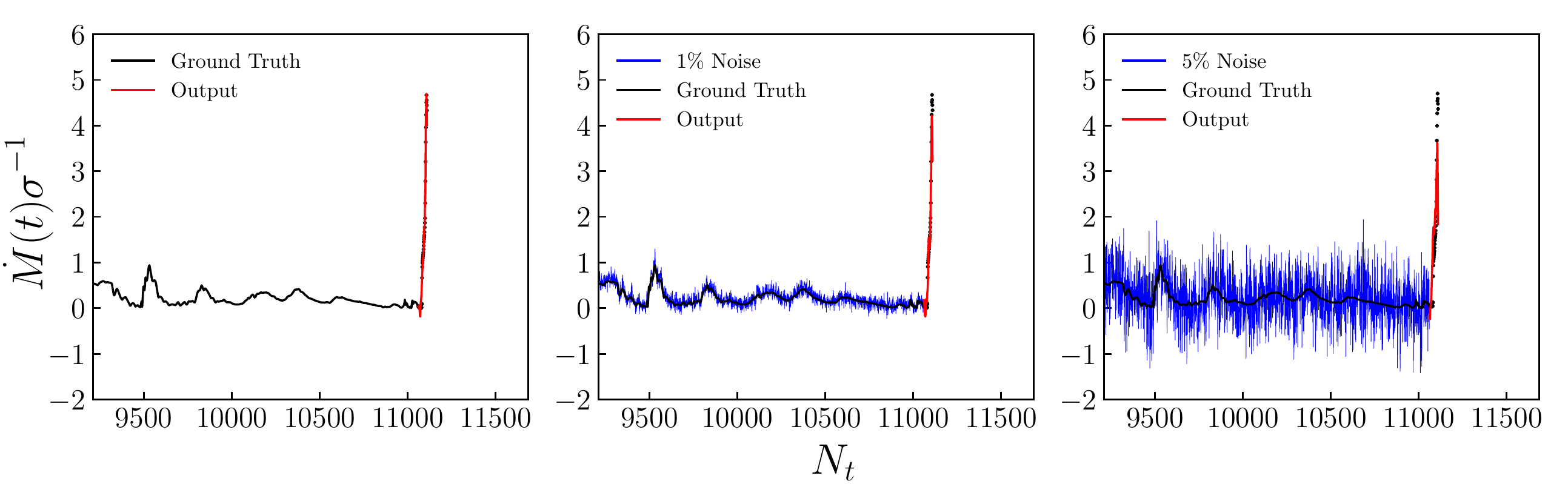}
  \caption{Episodic burst forecasts under noisy training conditions. Conditions range in severity from no noise (left panel) to 5\% noise addition (right panel). Here, the degree of noise is taken as a fraction (1 or 5 percent) of the maximum value of mass accretion rate. The performance results are given from left to right as $8.29 \times 10^{-3}$, $1.84\times 10^{-2}$, and $1.47\times 10^{-1}$, respectively. This demonstrates that the model trained under condition (1) is performing the best.} \label{burst_pred}
  \vspace{0.5cm}
  \includegraphics[width=\textwidth]{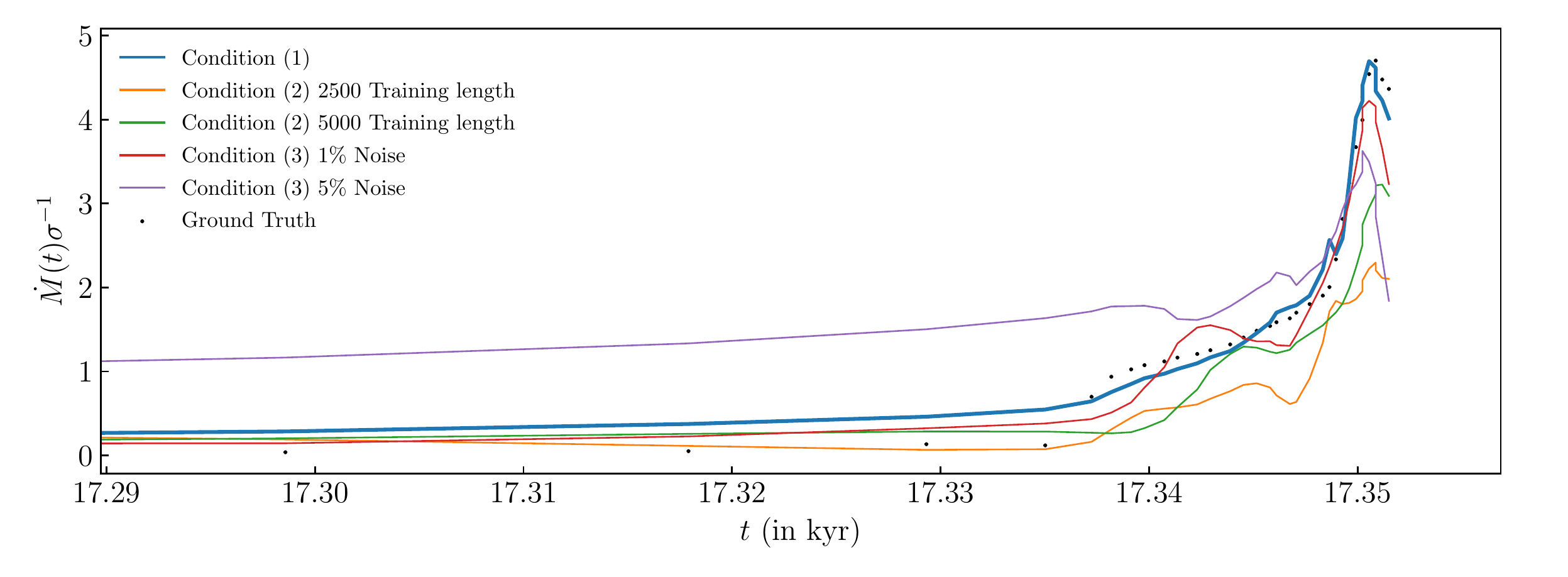}
  \vspace{-0.7cm}
  \caption{Comparison of burst forecasts across all training conditions (excluding 1000 training length for visualization purposes). Condition (1) gives the lowest NMSE ($8.29 \times 10^{-3}$), and is highlighted with increased line width.} \label{burst_pred_all}
\end{figure*}

 The simulation data across each model has demonstrated strong episodic behavior where the dynamics exhibit multiple high magnitude bursts that are driven largely by mass infall. The Opt-ESN has shown strong forecasting capabilities on segments of the data where episodic bursts have not occurred. An interesting hypothesis to test is whether the neural network can adequately resolve the occurrence of a burst over a short time interval. To do so, we follow the same protocol for data preparation and hyperparameter selection. We isolate the occurrence of the first large burst in Model 27 and produce forecasts over 50 time steps ($\approx 0.31$ kyr). To reflect some of the challenges in modeling observational data, we train the Opt-ESN under various conditions:
\begin{enumerate}
    \item Training with 10000 data points.
    \item Training with 5000, 2500 and 1000 data points.
    \item Training with 10000 data points with 1\% and 5\% noise added.
\end{enumerate}
These conditions additionally serve to assess the stability in the model's predictive power. That is, under nonideal training conditions, we aim to demonstrate that the network can still forecast meaningful outputs. Our training conditions vary in severity including scenarios with added noise and lack of data availability. We enumerate the training conditions below. Training condition (1) represents an ideal scenario with no noise and the entire dataset and has an NMSE of $8.29 \times 10^{-3}$. Condition (2) varies the training length, and condition (3) includes a moderate and high degree of noise in the data. The Opt-ESN performance and summary for each condition is given in Table \ref{tab::ep_hpars} and demonstrated in Figure \ref{burst_pred}. It is evident that the Opt-ESN demonstrates an ability to resolve the presence of an episodic burst with noisy training data. With increased noise the magnitude suffers and the NMSE increases relative to condition (1)  by $\approx 37\% \,\rm{and} \, 88\% $ when we included 1\% and 5\% noise, respectively. Condition (2) gives insight into the importance of data availability with respect to model performance. In this scenario, compared to condition (1), the NMSE increases by $\approx 10 \%$ and $  40 \%$  when the training length is reduced to 5000 and 2500 data points, respectively. 
\section{Discussion}
\label{discussion}
As we enter a new era in time domain astronomy, leveraging robust predictive models to make meaningful data inference is increasingly valuable. Neural networks became an attractive class of algorithms that can be used to model nonlinear data due to the universal approximation theorem \citep{cybenko1989approximation,schafer2006recurrent}. In particular, the reservoir computing framework that governs the ESN is ideal in modeling time series that possess chaotic (nonlinear) temporal structure. We used the Opt-ESN model to predict/forecast the protostellar mass accretion rate  by training it on simulated hydrodynamical time series data. To avoid nonstationary dynamics, first-order differencing was used on the deterministic components of the data. 
Our methodology has generated robust and accurate outputs over several sets of chaotic data (i.e., having a high degree of nonlinear temporal dynamics). 
\subsection{Short-term Burst Predictions and Data Availability}
\label{burst_pred_sec}
In Section \ref{burstsec}, we utilized the Opt-ESN to predict an episodic burst under various conditions. 
 Our findings demonstrate that the main driver in model degradation is a lack of data. Even with a relatively high degree of noise, the network was able to resolve the presence of a burst within a 50 time step forecast. The Opt-ESN forecasts are aimed to provide a introductory framework as to how ML based models can be used to make inference on burst occurrences and forecast mass accretion. This can possibly help in establishing any potential relation between FUor and EXor phenomena. In practice, however, having thousands of years of observational data on a single object is not feasible. We see in Figure \ref{burst_pred_all} that the model maintains relatively strong predictive performance in scenarios with training lengths as low as 2500 data points ($0.3092$ kyr). However, below this number, we see a significant decrease in model performance. Therefore, addressing data availability becomes a critical component in developing a robust neural network framework. In these scenarios, we propose two approaches:
\begin{enumerate}
    \item Simulating synthetic data. \\
    As demonstrated in our approach, we can leverage simulation data to train neural networks in conjunction with observational data. Here, the simulated data would be integrated with the observational data as part of the training/validating segments of the model. The simulated data however should reflect the physical properties of the observed object and should include several occurrences of episodic bursts to assure proper coverage in the possible dynamics.
    
    \item Aggregating data from similar objects. \\
    An alternative to generating synthetic data can be to aggregate data across several observed objects. That is, the neural network is trained across multiple similar objects. An advantage in this approach is that no portion in the training data is synthetic and as such, the forecasts will reflect the true dynamics of the observations. This approach can alleviate some of the data availability issues in practice, however  it is crucial that each of the observed objects reflect the physical properties of the entire system being modeled.
\end{enumerate}

\subsection{Effective Forecast Horizon}
Long forecast horizons in any statistical model can be challenging. This is due to the recursive dependence that is common in time series. That is, models typically will recycle outputs as future inputs in any out-of-sample forecasts. As such, the error in any given point estimate will propagate to future estimates, which limits the predictability of a given model. In the context of chaotic time series, the Lyapunov time becomes a natural scale on which to characterize an effective forecast horizon. That is, the time length on which one can achieve sufficiently reliable forecasts. Typically, data sets that exhibit more chaos in temporal dynamics will have a larger maximum Lyapunov exponent, which effectively limits the real time range in predictive power. \cite{pathak2018model} utilized the reservoir computing framework in predicting spatio-temporally chaotic systems for the Kuramoto-Sivashinsky equation. Their scheme achieves low prediction error for roughly 8 Lyapunov times. As such, for comparative purposes we used $\Lambda t_{\text{max}}= 8$ as a benchmark. Our Opt-ESN forecasts achieved low NMSE beyond the 8 Lyapunov time benchmark in each of the six simulation models (summarized in Table \ref{tab::per_summ}). Beyond this point, it becomes more likely that the forecasts become unreliable. In future work however, we may look into quantifying an asymptotic upper limit on the model's effective forecast horizon in terms of the Lyapunov time.
\\
\section{Conclusion}
\label{conclusion}
We introduced the Opt-ESN model to forecast the mass accretion in protostellar disk evolution. This model exploits the stochastic nature of echo state networks and introduces its use in time-domain astronomy. We applied our model to a series of synthetic mass accretion data sets simulated by solving the hydrodynamical equations for a protostellar disk. The model achieved predictions with a low normalized mean square error (NMSE) ($\sim 10^{-5}$ to $10^{-3}$) for forecasts ranging between 0.099 to 3.793 kyr. Additionally, the model successfully resolved the occurrence of an episodic burst with low NMSE when we added 1\% and 5\% of noise to the data. 
However, our findings also suggest that the model is not immune to degradation under scenarios of data limitation.

 


Our implementation demonstrates the predictive capabilities of Opt-ESN when applied to time series data. As we transition into a new era of time domain astronomy, understanding and developing robust statistical time series models is becoming increasingly important. 
Importantly, the scientific return of our work goes much beyond its application to observations in the optical domain. There may be synergies with observations made in the radio domain (e.g., fast radio bursts) with facilities like the Canadian Hydrogen Intensity Mapping Experiment (CHIME) and the Australian Square Kilometer Array Pathfinder (ASKAP). Likewise in the gravitational-wave arena, Opt-ESN can play a crucial role in detecting/denoising black hole and neutron star merger signals observed from facilities like the Laser Interferometer Gravitational-Wave Observatory (LIGO) and Virgo interferometer as well as the upcoming Kamioka Gravitational Wave Detector (KAGRA) and LIGO India. \\

\section*{acknowledgments}
 S.B. is supported by a Discovery Grant from NSERC. S.A. is supported by the NASA Postdoctoral Program (NPP). S.A. acknowledges that a portion of this research was carried out at the Jet Propulsion Laboratory, California Institute of Technology, under a contract with the National Aeronautics and Space Administration (80NM0018D0004). E.I.V. acknowledges support by the Ministry of Science and Higher Education of the Russian Federation (State assignment in the field of scientific activity 2023, GZ0110/23-10-IF).

\newpage
\appendix
Supplemental material is provided here for reference. This includes some background discussion on stationarity, as well as additional figures and tables that summarize particular sets of results. \\
\vspace{0.5cm}
\subsection{Stationarity}
A process is considered to be stationary if the underlying distribution is constant over time \citep{brockwell2002introduction, hamilton2020time}. This effectively translates to having constant values of its first four moments. Alternatively, we define a time series to be covariance-stationary if it is only constant in its first and second moments. More formally, let ${X_t}$ be a time series with $\mathbb{E}(X_t^2) < \infty$. The first and second moments are the mean and covariance functions, respectively:
\begin{align}
    \mu_X(t) &= \mathbb{E}(X_t), \\
    \Gamma_X(r,s) &= \text{Cov}(X_r, X_s), \quad \forall r,s \, .
\end{align}
Thus, the process ${X_t}$ is said to be covariance stationary if $\mu_X(t)$ is independent of $t$ and $\Gamma_X(t+h,t)$ is independent of $t$ for all $h$. Furthermore, the autocovariance function must be even and nonnegative definite \citep{brockwell2002introduction}. That is, for a real-valued vector $\textbf{A}$ having components $a_i$, we have
\begin{align}
\sum_{i,j} a_i \Gamma_X(i-j) a_j \geq 0 \, .
\end{align}
Additionally, a stationary process will have the roots of its characteristic equation lie inside the unit circle \citep{patterson2011unit}. That is, if the underlying process has a unit root $\geq 1$, then it is nonstationary. Assume the variable ${X_t}$ can be written as a $p^{\text{th}}$ order autoregressive process:
\begin{align}
    X_t = \alpha_1 X_{t-1} + \alpha_2 X_{t-2} + \dots + \alpha_p X_{t-p} + \epsilon_t \, ,
\end{align}
where the innovations $\epsilon_t$ are uncorrelated with mean zero and constant variance. If the characteristic equation
\begin{align}
    \lambda^p - \alpha_1 \lambda^{p-1} - \alpha_2 \lambda^{p-2} - \dots - \alpha_p =0
\end{align}
has roots $\lambda \geq 1$ of multiplicity $m$, then the process is nonstationary with integration order $m$, denoted $I(m)$. There are several approaches to assessing whether a time series is stationary or not. The augmented Dickey-Fuller (ADF) test is among a popular set of unit root tests for time series data. It involves testing the null hypothesis that a unit root is present in the underlying process, making the time series nonstationary \citep{patterson2011unit}. The ADF test assumes the underlying process can be modeled by
\begin{align}
    \Delta X_t = \alpha_0 + \alpha_1 t + \rho X_{t-1} + \gamma_1 \Delta X_{t-1} + \dots + \gamma_{p-1} \Delta X_{p-t+1} + \epsilon_t \, .
\end{align}
The process would have a unit root if $\rho = 1$ or alternatively be considered stationary if $\rho < 1$. As such, the test is carried out under the null hypothesis that $\rho = 1$ against the alternative that $\rho < 1$, where the test statistic $\text{DF}_{\rho}$ is given as
\begin{align}
\text{DF}_{\rho} = \frac{\hat{\rho}}{\text{SE}(\hat{\rho})}\, .
\end{align}
Here, $\text{SE}(\hat{\rho})$ is the standard error and the value $\text{DF}_{\rho}$ is compared to the respective critical value in the Dickey-Fuller distribution. In our implementation, we take stationarity to mean covariance-stationary and utilize the ADF test at 5\% significance with $p = 12\times(N_t/100)^{1/4}$ (default setting in Python), where $N_t$ is the number of observations.
\newpage
\vspace{0.5cm}
\subsection{Opt-ESN Model Outputs}
\begin{figure*}[h]
\centering
\hspace{-0.8cm}
\includegraphics[width=0.93\textwidth]{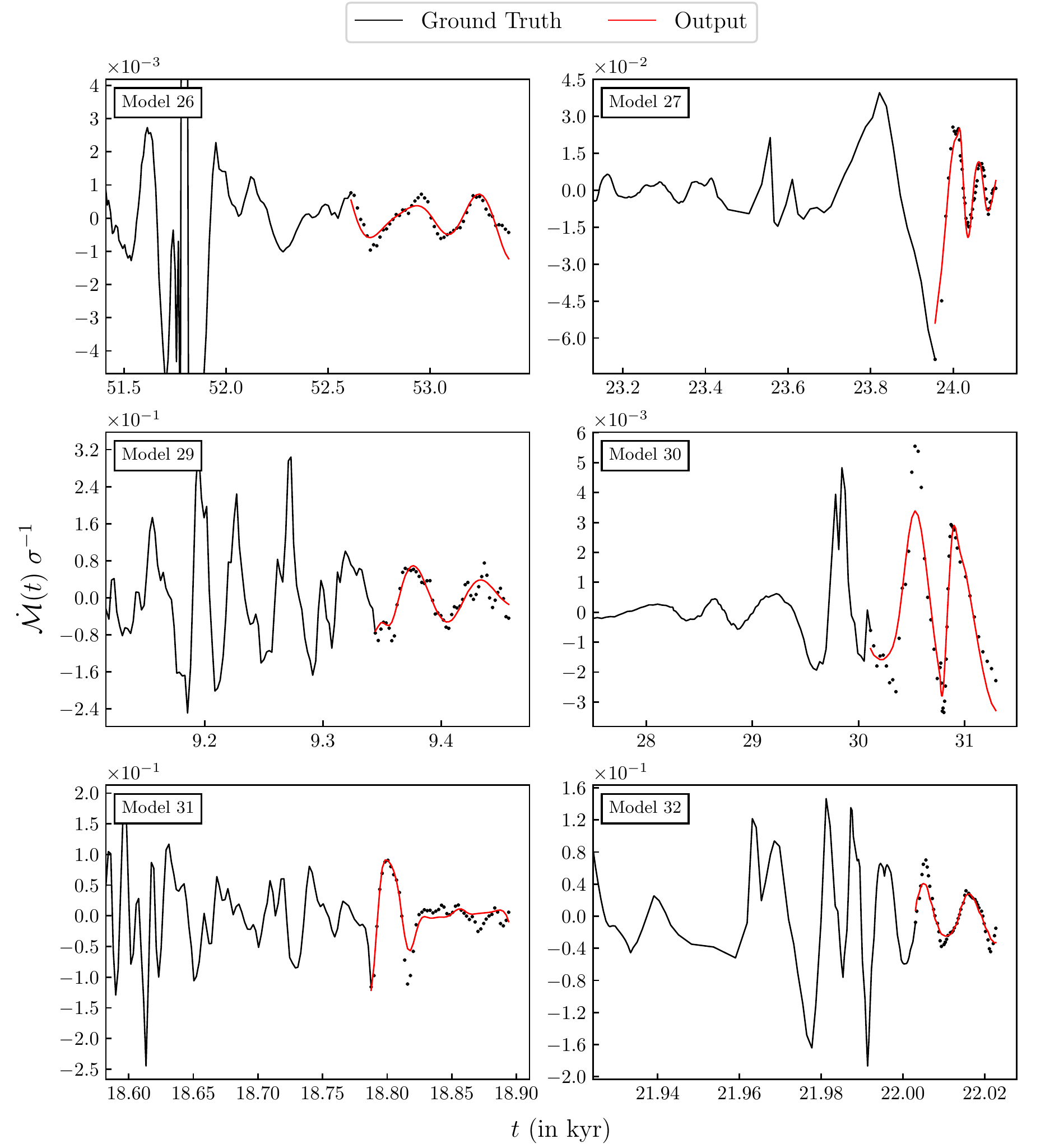}
  \caption{The Opt-ESN prediction on the fluctuating component of the simulation data. The solid black line is representative of the simulation data used in training/validation, the dotted black is the out-of-time testing segment of the data and the red line is the network's prediction.} \label{mod2732f}
\end{figure*}

 Figure~\ref{mod2732f} demonstrates the Opt-ESN prediction of the fluctuating component of the simulation data.

\begin{figure*}[h]
\hspace*{-1cm}
\centering
  \includegraphics[width=0.95\textwidth]{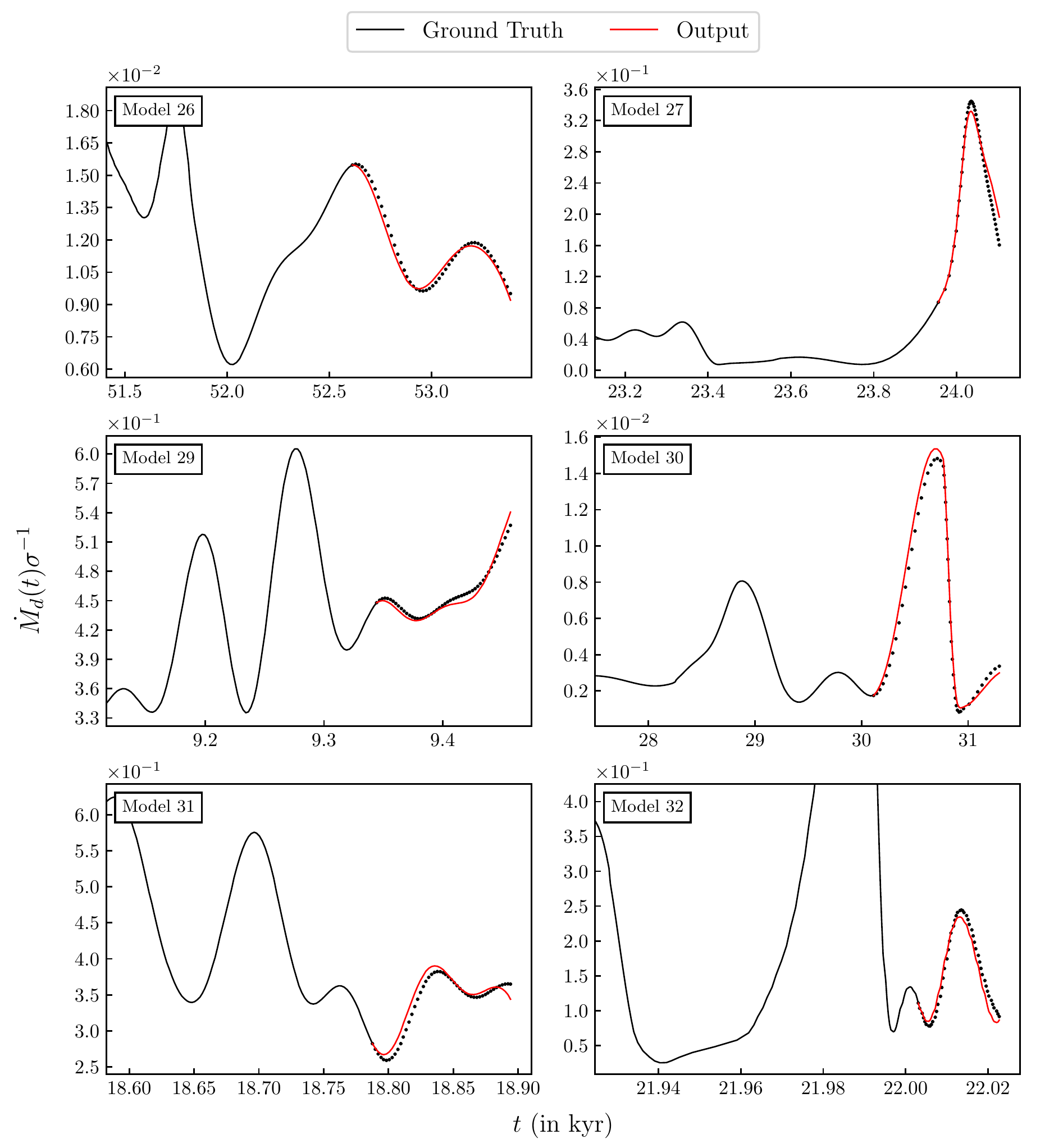}
  \caption{The Opt-ESN prediction on the deterministic component of the simulation data. The solid black line is representative of the simulation data used in training/validation, the dotted black is the out-of-time testing segment of the data and the red line is the network's prediction.} \label{mod2732d}
\end{figure*}

\newpage
Figure~\ref{mod2732d} demonstrates the Opt-ESN prediction of the deterministic component of the simulation data.

\newpage
\subsection{Performance Assessment} 
The tables below summarize the model performance metrics. Table \ref{tab::per_summ} contains the hyperparameters and forecast performances for each model component (fluctuating, deterministic, and overall). Furthermore, Table \ref{tab::ep_hpars} demonstrates the performance metrics on forecasting the episodic burst under each of the three training conditions. 

\begin{table*}[h]
\begin{center}
\caption{Opt-ESN hyperparameter settings and performance summary over each simulation model.} 
\begin{tabular}{clccccccccc}
\toprule\toprule 
    \multirow{2}{*}{\textbf{Model}} & 
    \multirow{2}{*}{\textbf{Component}} & 
    \multicolumn{4}{c}{\textbf{Hyperparameters}} &
    \multirow{2}{*}{ {$\Lambda$} } &
    \multirow{2}{*}{ {$\Lambda \cdot N_t$} } &
    \multirow{2}{*}{ {$t$ (kyr) } } &
    \multicolumn{2}{c}{\textbf{NMSE}}\\
    \cline{3-6}
    \cline{10-11}
    && {$\varrho$} & {$c_r$} & {$\alpha$} & {$\rho$} 
    &&&& Validation & Out-of-Time  \\
    \midrule
      & Fluctuating & 1.0000 & 0.1000 & 0.2000 & 0.5000 &   &   & & $1.00\mathrm{e}{-4}$ & $4.05\mathrm{e}{-5}$ \\
    Model 26 & Deterministic & 0.1000 & 0.1000 & 0.2000 & 0.7000 & 0.208 & 10.41 & 1.978 & $1.33\mathrm{e}{-5}$ & $1.05\mathrm{e}{-5}$ \\
      & Overall &   &   &   &   &   &   & & $5.11\mathrm{e}{-5}$ & $1.86\mathrm{e}{-5}$ \\
    \midrule
      & Fluctuating & 1.0000 & 0.1000 & 0.9000 & 0.9900 &   &   & & $3.75\mathrm{e}{-4}$ & $2.91\mathrm{e}{-4}$ \\
    Model 27 & Deterministic & 0.1886 & 0.1000 & 0.4655 & 0.1957 & 0.201 & 10.07 & 0.975 & $3.15\mathrm{e}{-4}$ & $1.01\mathrm{e}{-3}$ \\
      & Overall &   &   &   &   &   &   & & $3.84\mathrm{e}{-4}$ & $8.15\mathrm{e}{-4}$ \\
    \midrule
      & Fluctuating & 0.0769 & 0.0100 & 0.1117 & 0.6774 &   &   & & $1.68\mathrm{e}{-3}$ & $2.32\mathrm{e}{-3}$ \\
    Model 29 & Deterministic & 0.5000 & 0.1000 & 0.2000 & 0.5000 & 0.184 & 9.22 & 0.341 & $1.99\mathrm{e}{-4}$ & $2.94\mathrm{e}{-4}$ \\
      & Overall &   &   &   &   &   &   & & $1.34\mathrm{e}{-3}$ & $2.28\mathrm{e}{-3}$ \\
    \midrule
      & Fluctuating & 0.0298 & 0.0100 & 0.2680 & 0.3306 &   &   & & $1.78\mathrm{e}{-4}$ & $7.97\mathrm{e}{-5}$ \\
    Model 30 & Deterministic & 0.0783 & 0.0100 & 0.1970 & 0.2960 & 0.162 & 8.12 & 3.793 & $1.53\mathrm{e}{-5}$ & $2.15\mathrm{e}{-5}$ \\
      & Overall &   &   &   &   &   &   & & $1.83\mathrm{e}{-4}$ & $5.93\mathrm{e}{-5}$ \\
    \midrule
      & Fluctuating & 0.5000 & 0.1000 & 0.5000 & 0.8500 &   &   & & $1.06\mathrm{e}{-3}$ & $1.23\mathrm{e}{-3}$ \\
    Model 31 & Deterministic & 0.5000 & 0.1000 & 0.5000 & 0.2000 & 0.199 & 9.94 & 0.312 & $1.58\mathrm{e}{-4}$ & $8.95\mathrm{e}{-4}$ \\
      & Overall &   &   &   &   &   &   & & $1.15\mathrm{e}{-3}$ & $1.91\mathrm{e}{-3}$ \\
    \midrule
      & Fluctuating & 0.0673 & 0.0100 & 0.3224 & 0.2933 &   &   & & $7.64\mathrm{e}{-4}$ & $9.50\mathrm{e}{-4}$ \\
    Model 32 & Deterministic & 0.0481 & 0.0100 & 0.1590 & 0.3532 & 0.192 & 9.61 & 0.099 & $1.02\mathrm{e}{-4}$ & $9.98\mathrm{e}{-4}$ \\
      & Overall &   &   &   &   &   &   & & $2.85\mathrm{e}{-4}$ & $1.18\mathrm{e}{-3}$ \\
    \hline \hline
\end{tabular} \label{tab::per_summ}
\end{center}
\end{table*}
\begin{table*}[h]
\begin{center}
\caption{Opt-ESN model settings used in episodic burst predictions.} 
\vspace{0.2cm}
\begin{tabular}{clccccccc}
\toprule\toprule
\multirow{2}{*}{\textbf{Condition}} & \multirow{2}{*}{\textbf{Component}} &\multicolumn{4}{c}{\textbf{Hyperparameters}} & \multirow{2}{*}{\textbf{NMSE}} & 
\multirow{2}{*}{\textbf{Training length}} &
\multirow{2}{*}{\textbf{Level of noise}}
\\
\cline{3-6}
  &&{$\varrho$} & {$c_r$} & {$\alpha$} & {$\rho$}& \\
\midrule
    & Fluctuating   & $0.6661$ & $0.1000$ & $0.1836$ & $0.6394$ & $1.64$e$-2$     \\
(1) & Deterministic & $0.8185$ & $0.1000$ & $0.2583$ & $0.2953$ & $2.46$e$-4$ & 10000 & 0\% \\
    & Overall       &   &    &   &   & $8.29$e$-3$     \\
\midrule
     & Fluctuating   & 0.0390 & 0.1000  & 0.1535 & 0.4175 & $2.24$e$-2$    \\
     & Deterministic & 0.2091 & 0.1000  & 0.3273 & 0.5917 & $1.81$e$-3$ & 5000 & 0\%\\
     & Overall       &   &    &   &   & $1.07$e$-2$     \\
     \cline{2-9}
     & Fluctuating   & 0.0474 & 0.1000 & 0.4446 & 0.1319 & $5.16$e$-2$      \\
(2)  & Deterministic & 0.6935 & 0.1000  & 0.7389 & 0.3642 & $3.10$e$-3$ & 2500 & 0\%\\
     & Overall       &   &    &   &   & $2.25$e$-2$     \\
     \cline{2-9}
     & Fluctuating   & 0.7489 & 0.1000  & 0.0102 & 0.9848 & $1.14$e$+5$      \\
     & Deterministic & 0.0184 & 0.1000  & 0.4612 & 0.1193 & $1.74$e$+0$ & 1000 & 0\%\\
     & Overall       &   &    &   &   & $6.38$e$+4$ 
     \\
\midrule
     & Fluctuating   & $0.3026$ & $0.1000$  & $0.1755$ & $0.8945$ & $2.58$e$-2$   \\
     & Deterministic & $0.6681$ & $0.1000$  & $0.4042$ & $0.2133$ & $1.27$e$-3$ & 10000 & 1\% \\
(3) & Overall       &   &    &   &   & $1.84$e$-2$     \\
     \cline{2-9}
     & Fluctuating   & $1.0000$ & $0.1000$  & $0.3000$ & $0.9000$ & $6.81$e$-2$     \\
     & Deterministic & $0.7527$ & $0.1000$  & $0.0296$ & $0.8296$ & $1.36$e$-1$ & 10000 & 5\% \\
     & Overall       &   &    &   &   & $1.47$e$-1$ \\
\hline \hline
\end{tabular}\label{tab::ep_hpars}
\end{center}
\end{table*}
\newpage
\bibliography{bibfile}
\bibliographystyle{aasjournal}

\end{document}